\documentclass[journal]{IEEEtran}

\usepackage{xspace}
\makeatletter
\DeclareRobustCommand\onedot{\futurelet\@let@token\@onedot}
\def\@onedot{\ifx\@let@token.\else.\null\fi\xspace}

\def\ie{i.e\onedot}

\makeatother

\usepackage{amsthm,amsmath,amssymb,amsfonts}
\usepackage[caption=false, font=footnotesize]{subfig}

\usepackage{amsmath}
\usepackage{graphicx}
\usepackage{algorithm}
\usepackage{algpseudocode}
\usepackage{bm}
\usepackage[inline]{enumitem}

\newtheorem{definition}{Definition}

\usepackage{commath}
\usepackage{hyperref} 
\usepackage[capitalize]{cleveref}
\Crefname{section}{Section}{Sections}
\Crefname{table}{Table}{Tables}
\Crefname{algorithm}{Algorithm}{Algorithms}
\Crefname{theorem}{Theorem}{Theorems}
\Crefname{lemma}{Lemma}{Lemmas}
\Crefname{definition}{Definition}{Definitions}
\Crefname{proposition}{Proposition}{Propositions}
\usepackage{booktabs}

\begin{document}

\title{Towards High-Performance Network Coding: FPGA Acceleration With Bounded-value Generators}

\author{Jiaxin~Qing,
        Philip~H.~W.~Leong,~\IEEEmembership{Senior~Member,~IEEE,}
        Kin-Hong~Lee,
        and~Raymond~W.~Yeung,~\IEEEmembership{Fellow,~IEEE}%
        
\thanks{%
    J.~Qing is with the Department of Information Engineering, The Chinese University of Hong Kong, Hong Kong SAR (email: \href{mailto:jqing@ie.cuhk.edu.hk}{jqing@ie.cuhk.edu.hk}).
}%
\thanks{P.~H.~W.~Leong is with the School of Electrical and Information Engineering, The University of Sydney, Australia (email: \href{mailto:philip.leong@sydney.edu.au}{philip.leong@sydney.edu.au}).}%
\thanks{Kin-Hong~Lee is with the Department of Computer Science Engineering, The Chinese University of Hong Kong, Hong Kong SAR (email: \href{mailto:khlee@cse.cuhk.edu.hk}{khlee@cse.cuhk.edu.hk}).}
\thanks{R.~W.~Yeung is with the Department of Information Engineering, The Chinese University of Hong Kong, Hong Kong SAR. R.~W.~Yeung is also with the Institute of Network Coding, The Chinese University of Hong Kong, and he is also a Principal Investigator of the Centre for Perceptual and Interactive Intelligence (CPII) Limited (email: \href{mailto:whyeung@ie.cuhk.edu.hk}{whyeung@ie.cuhk.edu.hk}). The work of R. W. Yeung was supported in part by a fellowship award from the Research Grants Council of the Hong Kong Special Administrative Region, China under Grant CUHK SRFS2223-4S03.}
}

\maketitle

\begin{abstract}
Network coding enhances performance in network communications and distributed storage by increasing throughput and robustness while reducing latency. Batched Sparse (BATS) codes are a class of capacity-achieving network codes, but their practical applications are hindered by their structure, computational intensity, and power demands of finite field operations.
Most literature focuses on algorithmic-level techniques to improve coding efficiency. Optimization with an algorithm/hardware co-designing approach has long been neglected.
Leveraging the unique structure of BATS codes, we first present CS-BATS, a hardware-friendly variant. Next we propose a simple but effective bounded-value generator, to reduce the size of a finite field multiplier by up to 70\%.  Finally, we report on a scalable and resource-efficient FPGA-based network coding accelerator that achieves a throughput of 27~Gbps, a speedup of more than 300 over software. 
      
\end{abstract}


\section{Introduction}
\label{sec:intro}
\IEEEPARstart{I}{n} the modern era of information exchange and data-driven communication, the efficient and secure transfer of digital content lies at the heart of our interconnected world. 
Traditional approaches to data transmission consider data packets as passive entities that can only be routed and forwarded in the network. 
Network coding~\cite{netcod} can be seen as the digital equivalent of an alchemical transformation that allows intermediate nodes in a network to intelligently combine, restructure, and encode incoming data packets before forwarding them toward their intended destinations~\cite{l_necod}.
This transformative approach not only enhances data throughput and network robustness but also promises significant gains in terms of resource utilization and security~\cite{s_netcod, rand_netcod, good_netcod}.
The applications of network coding span a wide spectrum of network types, including wired and wireless networks~\cite{wireless_netcod1, wireless_netcod2, phy_netcod}, peer-to-peer systems~\cite{rand_netcod}, content distribution networks~\cite{content_netcod, coded_cache1} and distributed storage~\cite{dist_store}. 

While network coding holds the promise of revolutionizing data transmission and network efficiency, its applications in real-world networks are challenging.
For example, \cite{prac_necod} concluded that the performance of network coding was significantly limited by its computation intensity, coding latency, and buffering latency. To solve these issues, many advanced network coding schemes have been proposed. Most of these works focus on optimizing the coding scheme from an algorithm level, such as reducing coding complexity, buffering complexity~\cite{netcod4, netcod5, prac_necod2, spare_code}, or improving network control methods to reduce computation requirement~\cite{imp1, imp2}. 
Among them, Batched Sparse Codes (BATS codes)~\cite{bats-tit} is the most practical code, with provable capacity-achieving capability under various network topologies.  
However, its applications are still hindered by its random structure and high computation intensity, especially under the increasing demands for higher data rates and lower power consumption. For example, 5G wireless networks require a peak data rate of 20~Gbps~\cite{5g-standard}, while an optimized BATS implementation on an Intel I7-10700 CPU with C++ can only achieve a data rate of around 100~Mbps~\cite{tvt}.  

Hardware has evolved significantly over the past decades, and new technologies, particularly high-bandwidth memory (HBM), have profoundly influenced computer architectures~\cite{he-com, fpga-hbm1, fpga-hbm2}. These improvements in hardware and computation power have fundamentally altered the landscape of many fields, including neural networks, wireless communications, and science.  
Thus, we raise the question: \textit{Is it time to revisit the practicality of network coding?}

We shed light on this problem with a highly performant and scalable hardware design for the BATS code. 
Through an algorithm-hardware co-designing scheme, we re-design the traditional random BATS code into a structured one, where input packets are accessed in a periodic pattern. 
By exploiting the HBM of AMD Alveo U50~\cite{u50} and carefully integrating the BATS code with the FPGA, we can easily achieve a throughput of more than 27~Gbps with a small logic and power consumption. 
The contributions are as follows. 
\begin{itemize}
\item This is the first work to quantitatively compare the hardware efficiency of conventional BATS codes with Cyclic-Shift BATS. 
\item We propose a scalable and adaptable primitive compute unit that achieves batch-level parallelism with lower resource consumption and higher performance than previous works.
\item We introduce bounded-value generators with theoretical and experimental guarantees, which reduce the size of a finite-field multiplier up to 70\% without impacting the coding performance.    
\item Our techniques are combined in an FPGA-based accelerator, which, to the best of our knowledge, is the highest-throughput capacity-achieving network coding accelerator reported. While reference~\cite{tvt} reported a throughput of 1.2 Gbps, this work achieves 27~Gbps on an AMD Alveo U50.
\end{itemize}

\section{Background}
The BATS code has been intensively studied in communication research, where it is adapted to various use cases~\cite{bats_unequal, bats_video, bats_v2v}. However, most literature does not consider the computer platform on which the BATS code is implemented. Hardware-software co-design, where the algorithms and hardware are jointly optimized, is crucial to match the increasing demands for throughput. Numerous examples can be found in the recent development of deep learning and communications research~\cite{flashattn, tpu, ldpc-survey}. 

The BATS code was firstly analyzed from a hardware perspective in \cite{tvt}, where the hardware performance upper bound is derived from a theoretical model based on the maximum possible number of finite field multipliers on the hardware. The paper also showed that the code construction, implemented as tiled matrix multiplication on an Intel Cyclone~V FPGA, can achieve a maximum throughput of 1.2~Gbps. Even though the FPGA implementation achieved a significant performance improvement compared with CPU-based implementation, the design is large and not scalable. It utilizes almost all logic resources on the FPGA to achieve that throughput.

\subsection{BATS Code}
\begin{figure}
    \centering
    \includegraphics[width=1\linewidth]{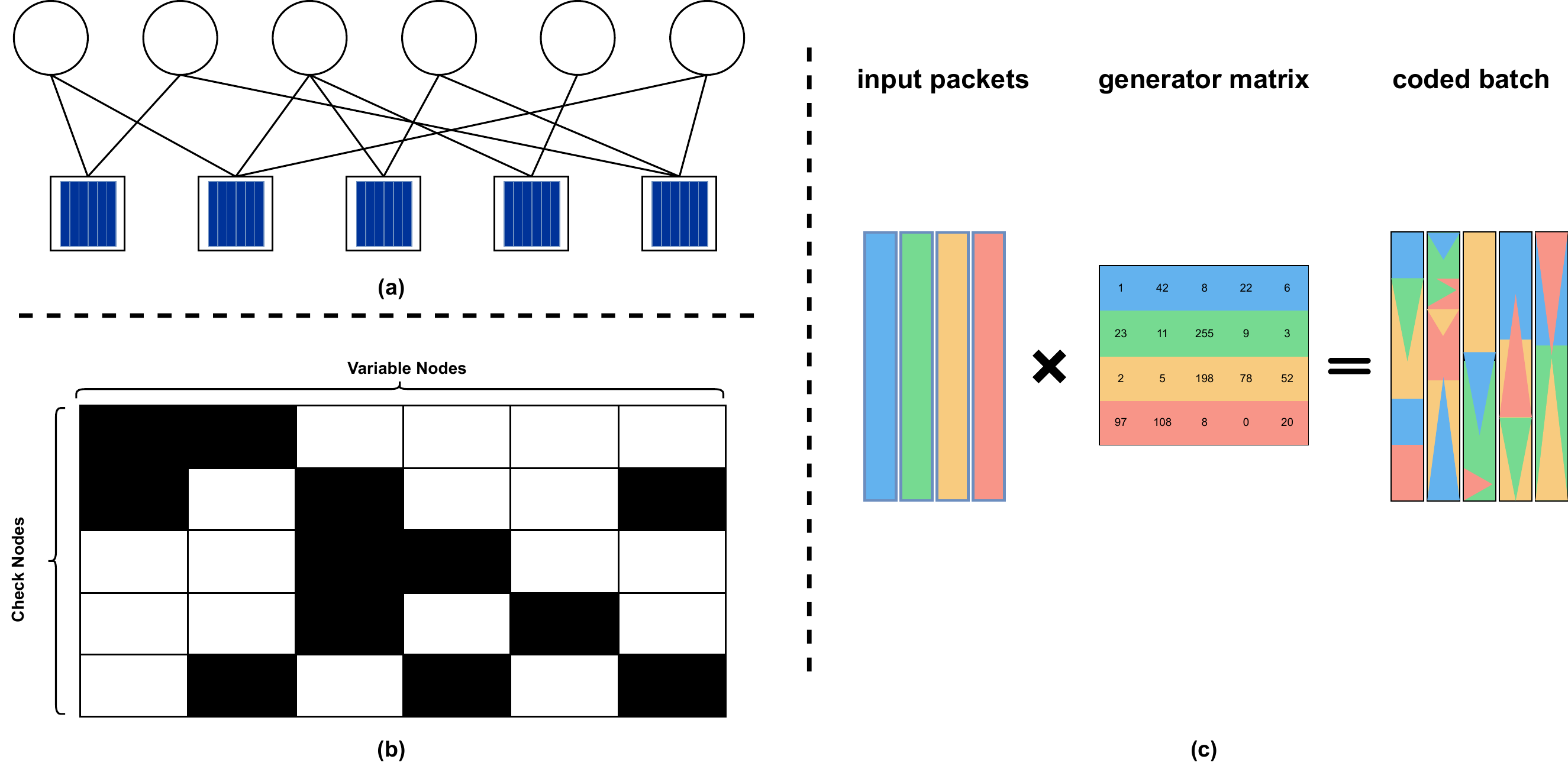}
    \caption{\textbf{Graphical description of the BATS code.} (a) Tanner graph representation. Circles are variable nodes, and squares are check nodes; (b) Adjacency matrix representation. Each darkened cell represents a connection;
    (c) Encoding process. Linear combinations mix information from input packets. }
    \label{fig:tanner}
\end{figure}

The BATS code is usually applied to a stream of data. Assuming that the data stream is of depth $F$ bits and the Galois field $GF(2^n)$ is used, the data stream can be divided into $K=F/(n*pk)$ packets, where each packet consists of $pk$ finite field elements, and each finite field element represents $n$ bits of data.
The BATS code encodes the $K$ input packets by taking linear combinations of different subsets of $K$, which the following linear system can describe,
\begin{equation}
    \bm{X}_i = \bm{B}_i\bm{G}_i,
    \label{eq:encoding}
\end{equation}
where $\bm{B}_i\in\mathbb{F}_q^{pk\times dg_i}$ is the $i$-th subset with $dg_i$ packets, and  $\bm{G}_i\in\mathbb{F}_q^{dg_i\times M}$ is the generator matrix that defines the linear combination of this subset. $\bm{X}_i\in\mathbb{F}_q^{pk\times M}$ is the $i$-th constructed batch, which consists of $M$ coded packets. 
The batches will be transmitted to the destination via lossy channels in a network. Thus, the number of generated batches $N$ depends on the packet loss rate during transmission. 

The construction of a BATS code can be visualized as the construction of a Tanner graph~\cite{tanner-graph}, where the variable nodes correspond to the input packets, and the check nodes correspond to generated batches. 
The BATS code can also be characterized using an adjacency matrix with rows representing the batches and columns representing the input packets, as shown in \cref{fig:tanner}.
The encoding can also be conceptualized as mixing information from selected input packets through linear combinations characterized by a generator matrix. 
At the intermediate nodes, packets will be linearly combined again. At the destination node, the input packets can be recovered by solving systems of linear equations. The two processes at the intermediate and destination nodes are called recoding and decoding, respectively. Here, we focus on the construction process. 
Readers are referred to \cite{bats-tit} for a detailed discussion on the recoding and decoding.

\subsection{Random BATS}
The choice of $dg_i$ is critical for the code's performance. The original BATS code relies on sampling $dg_i$ from a probabilistic distribution optimized for a high decoding rate~\cite{bats-tit, bats-finite}. After degree sampling, input packets will be randomly selected to construct a batch. 

This random structure leads to two issues for hardware implementation. Firstly, since the $dg_i$ is determined on the fly, the accelerator needs to handle the maximum degree, which is $K$ for all the input packets. In practice, $K=256$ is a frequently used value, and the optimized degree distribution will usually concentrate around the batch size $M$, where $M=16$ is usually used~\cite{bats-finite, tvt, bats-rl}. Thus, although this occurrence is low, logic resources must be reserved for buffering and computing 256 packets. 
Secondly, random addressing leads to low I/O efficiency. The selected packets and their sequence are unknown until runtime. This will usually cause failure in many I/O optimization techniques, such as bursting, coalescing, and pre-fetching. External memory I/O is most efficient when consecutive addresses are accessed, or the data is accessed with a certain pattern~\cite{hls_guide}. 

\subsection{Cyclic-Shift BATS (CS-BATS)}
\begin{figure}
    \centering
    \includegraphics[width=0.8\linewidth]{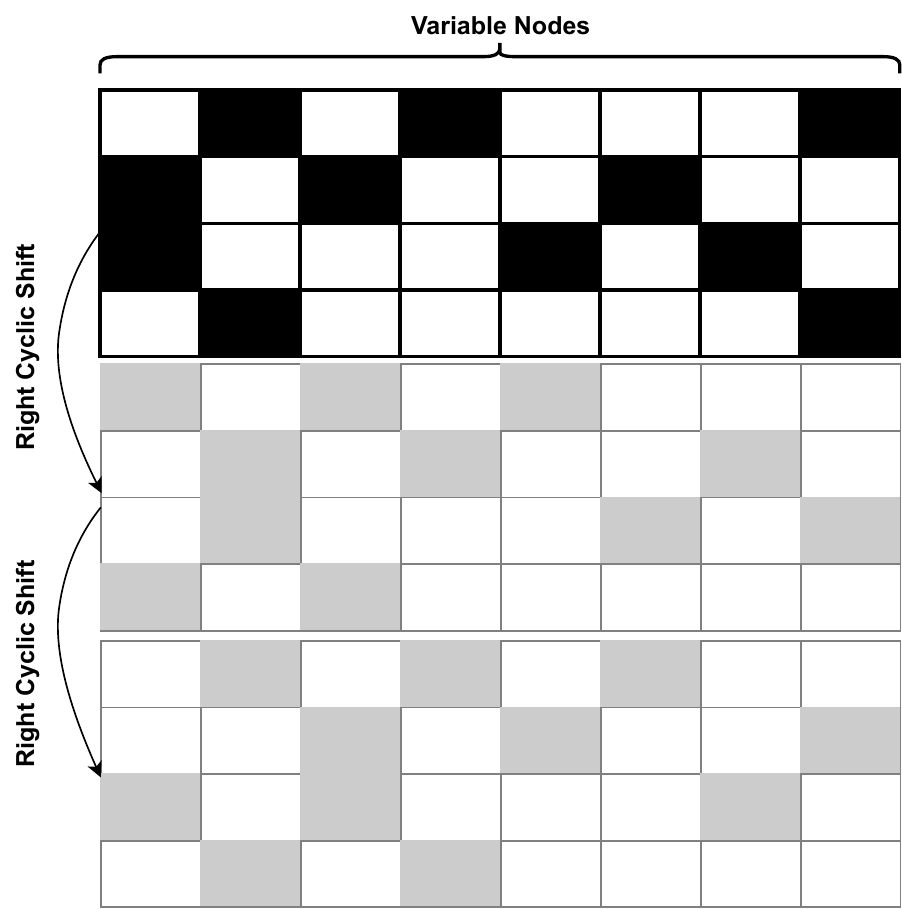}
    \caption{\textbf{CS-BATS construction from a base graph.} The base graph is a 4 by 8 adjacency matrix. Cyclic shifting is applied to each row of the base graph to construct new rows. }
    \label{fig:cs-bats}
\end{figure}

The CS-BATS changes the random construction into a structured construction~\cite{cs-bats}. 
Previous performance analysis implicitly or explicitly assumes check node independence~\cite{bats-tit, bats-finite, bats-tree}. However, the check nodes can be highly dependent, especially when $K$ is finite, and the Tanner graph is constructed randomly. It is shown that check node dependence has a significant negative impact on the code's performance~\cite{cs-bats}. The CS-BATS alleviates check node dependence through a structured construction. 
Even though CS-BATS demonstrates superior performance over the random BATS in software simulations~\cite{cs-bats}, its advantages in hardware design have not been well explored. 

\paragraph{Construction}
The CS-BATS constructs the Tanner graph from a small base graph, as shown in \cref{fig:cs-bats}. The base graph is characterized by an adjacency matrix of $m$ by $K$. To generate the $i$-th batch, the $(i \bmod{m})$-th row of the base graph will be cyclically shifted (either left or right) $\lfloor{i / m}\rfloor$ times.      

\paragraph{Complexity}
If the maximum degree of the base graph is $d_{max}$, we only need to reserve sufficient resources to buffer and compute $d_{max}$ packets rather than $256$ packets. According to \cite{cs-bats}, using a structured design, large degrees in the degree distribution can be discarded. Actually, the CS-BATS using a base graph with $d_{max}=27$ already has an improved decoding rate over the random BATS. 
Additionally, since the data access pattern is pre-defined, packets can be prefetched to maximize I/O utilization. Thus, we propose a hardware implementation based on the CS-BATS rather than the random BATS.

\section{BATS Compute Unit}
\begin{figure}
    \centering
    \includegraphics[width=0.6\linewidth]{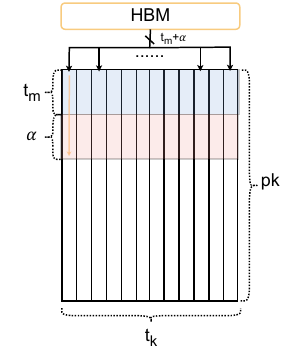}
    \caption{Decouple loading and computing. When loading a tile of $t_m\times t_k$ from matrix \textbf{B}, for each data transaction, we load $t_m+\alpha$ elements from its column. The $\alpha$ is chosen to fully utilize the memory bandwidth. }
    \label{fig:decouple}
\end{figure}

\begin{figure*}
    \centering
    \includegraphics[width=1\linewidth]{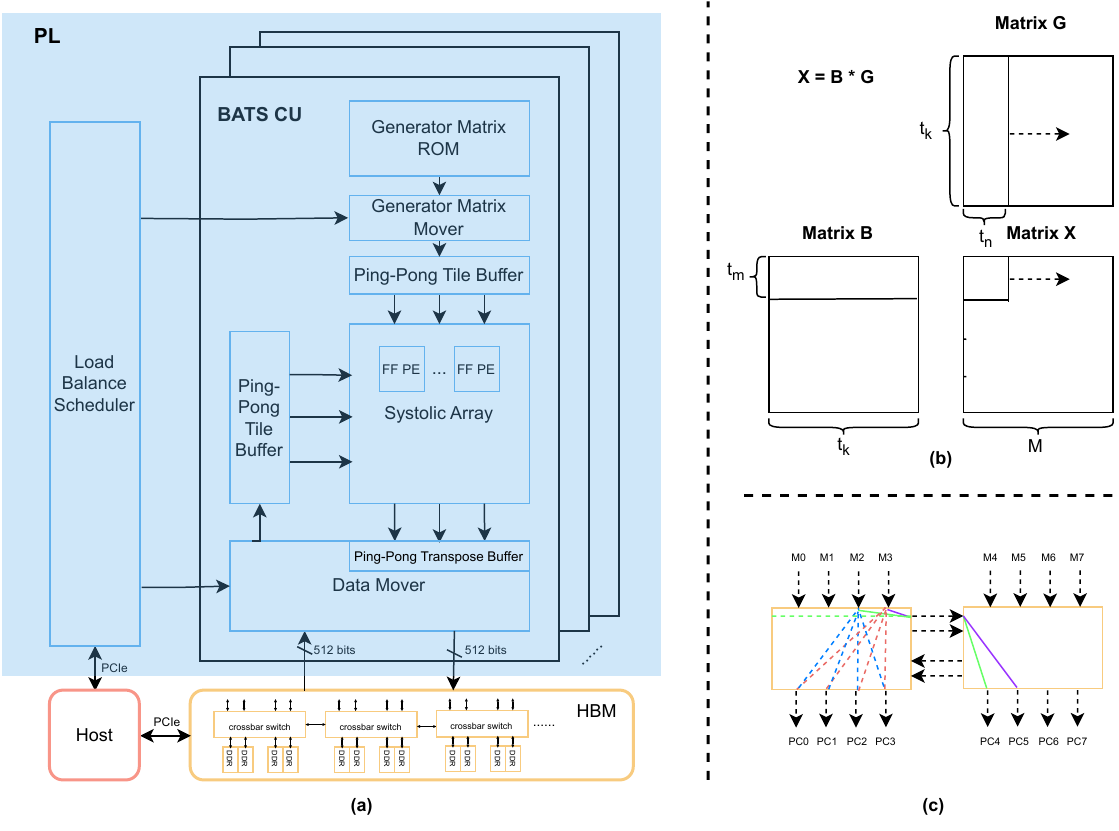}
    \caption[System diagram]{(a) Block diagram of the BATS accelerator system. FF PE: Finite field processing elements; BATS CU: BATS compute unit; The accelerator can have multiple BATS CUs. Each BATS CU accesses the HBM through an individual HBM pseudo channel (PC). With the Xilinx HLS Vitis flow, the maximum port width between the PL and the HBM is 512 bits. (b) Matrix tiling along the row, which reuses the tiles of matrix B. (c) The internal routing of the crossbar switch.}
    \label{fig:block}
\end{figure*}
The target application scenario considered is wireless communication, where an erasure channel is introduced by the network protocol stacks, such as TCP~\cite{bats-proto}. Given that the data to be encoded are typically generated or processed by other modules within these protocol stacks, we assume that the data are stored in an off-chip memory, accessible by the programmable logic (PL) via AXI interconnects.

To achieve a scalable design, we propose the BATS Compute Unit (BATS CU), which performs the fundamental task of BATS code construction, specifically batch generation. BATS CUs are designed to operate independently and in parallel, allowing for replication on the FPGA to meet the demands of higher throughput.

The block design of a BATS CU is depicted in \cref{fig:block}. It comprises a computation unit, two data-moving units (Data Mover and Generator Matrix Mover), a control and scheduling unit, and an on-chip ROM. The design of each unit will be discussed in detail in this section.

In essence, the construction of a BATS code involves finite field matrix multiplication, as described by \cref{eq:encoding}, where the matrix \( B_i \) is significantly larger than the matrix \( G_i \). The matrix \( \bm{B}_i \in \mathbb{F}_q^{pk \times dg_i} \) represents the selected packets, which typically range from a few kilobytes in size, depending on the degree and packet size. The matrix \( \bm{G}_i \in \mathbb{F}_q^{dg_i \times M} \) is a generator matrix, generally occupying a few hundred bytes. For clarity, the batch index \( i \) will be omitted in the subsequent discussion when there is no risk of ambiguity.

To efficiently implement code construction, matrix tiling is employed, dividing the matrix multiplication into the multiplication of several submatrices. This technique is a well-established approach in matrix multiplication, known for enhancing data reuse and enabling parallel computation. However, it must be carefully adapted to suit the specific requirements of our problem. In general, we identified the following challenges. 
\begin{itemize}
    \item Typically, matrix $\bm{B}$ is very tall and its size is large compared with matrix $\bm{G}$. How should these matrices be stored and loaded?
    \item The columns of $\bm{B}$ are usually not stored contiguously in the memory. What is the best tiling strategy to maximize data reuse and I/O efficiency?  
    \item Since the construction of each batch is independent of each other, how do we exploit batch-level parallelism? 
    \item The HBM has limited channels. How do we efficiently utilize the HBM channels to make the design more scalable? 
\end{itemize}

\subsection{Decouple Loading and Computing}

\begin{algorithm}[t]
    \caption{Decoupling of data access and tiling}\label{alg:data-access}
    \begin{algorithmic}[1]
        \State \textbf{Input} Off-chip memory port $\mathcal{M}$; Memory addresses $\{a_1, a_2, ..., a_{t_k}\}$; Empty tile buffer, $\hat{\bm{T}}$, of size $(t_m+\alpha)\times t_k$
        \State $\alpha=\mathrm{memory\_bandwidth} / \mathrm{element\_size} - t_m$
        \For{$k$ in \{1, 2, ..., $t_k$\}}
            \If{$\mathcal{M}[a_k:a_k+t_m]$ not in $\hat{\bm{T}}$}
                \State $\hat{\bm{T}}[:,k] \leftarrow \mathcal{M}[a_k:a_k+t_m+\alpha]$
            \EndIf
        \EndFor
        \State \textbf{Return} $\hat{\bm{T}}$
    \end{algorithmic}
\end{algorithm}

\subsubsection{Matrix tiling}
When implementing the CS-BATS, we can generate a fixed set of full-rank generator matrices $\{\bm{G}_1, \bm{G}_2, ..., \bm{G}_m\}$ according to the base graph and reuse them for all new batches. Thus, to balance I/O efficiency and resource consumption, the generator matrices can be stored in on-chip ROM, and the selected packets are tiled and accessed from an off-chip memory.
To reuse data in $\bm{B}$ and reduce off-chip memory accesses, the tiling is performed along the row of the matrix $\bm{X}$, as illustrated by \cref{fig:block}b, where the $t_k$ is equal to the number of columns of matrix $\bm{B}$.\footnote{In practice, $t_k$ will be the maximum base graph degree.} Matrix $\bm{B}$ tile consisting of $t_m\times t_k$ elements will be reused when computing the tiles of matrix $\bm{X}$ along its row.

At runtime, the host sends the input data address and the number of batches needed to the scheduler through PCIe. Then, the data mover reads a tile of size $t_m\times t_k$, namely, a submatrix of $\bm{B}$, from the off-chip memory. 
Denote $\bm{T}_{j}$ as the matrix $\bm{B}$ tile indexed by $j$, where $j\in[1, pk/t_m]$, and it is assumed that $t_m$ divides $pk$. 


\subsubsection{Maximize memory bandwidth utilization}
We will load the tile $\bm{T}_{j}$ along its column, as the elements of a column are from the same packet and stored contiguously in the memory. 
However, if we only read one column of data in one data transaction, the memory bandwidth utilization is limited by $t_m$, leading to a possible insufficient memory bandwidth usage. 
While we can increase $t_m$ to exploit the memory bandwidth fully, it will increase the resource consumption of the computation unit in return, as the systolic array in the computation unit is built according to the tile size. 

Thus, we decouple the tile size and the number of elements to read during one data transaction with \cref{alg:data-access}.
Specifically, when reading each tile column, instead of $t_m$ elements, $t_m+\alpha$ contiguous elements will be read and transferred into the tile buffer. 
The value of $\alpha$ can be either tailored according to specific resource constraints or used to fully exploit the memory bandwidth.
Notice that \cref{alg:data-access} and \cref{fig:decouple} describe the loading of one tile of $\bm{B}$, and the tile index $j$ is omitted.
In this way, the total number of data transactions can be reduced from $\frac{M * pk}{t_m}$ to $\frac{M * pk}{t_m+\alpha}$, which is derived by counting.  
The tile buffer is structured as a ping-pong buffer to overlap the data transfer with the computation.

A ping-pong buffer is also used to store tiles of $\bm{G}_i$. Upon receiving activation from the scheduler, the generator matrix mover reads tiles of size $t_k\times t_n$ from the ROM and stores the tile in the buffer. 

\subsection{Multi-level Parallelism and HBM}

\begin{algorithm}
    \caption{Output Ping-Pong buffering for a BATS CU}\label{alg:cu-ping-pong}
    \begin{algorithmic}[1]
        \State \textbf{Input} Computed tiles of size $t_m\times t_n$; Output buffers $b_0, b_1$, each of size $t_m\times M$
        \State $\#\mathrm{tiles\_in\_buffer}=M/t_n$
        \State $i=0$
        \For{tiles in all computed tiles}
            \State $b_i\leftarrow tiles$
            \If{$b_i$ is full}
                \If{$b_{i\oplus 1}$ has been transferred to off-chip memory}
                    \State send $b_i$ to paralleled data mover
                     \State $i\leftarrow i\oplus 1$
                \Else
                    \State stall
                \EndIf
            \EndIf
           
        \EndFor
    \end{algorithmic}
\end{algorithm}

\begin{algorithm}
    \caption{Output port sharing across BATS CUs}\label{alg:data-out}
    \begin{algorithmic}[1]
        \State \textbf{Input} Off-chip memory port $\mathcal{M}$; Output buffers of CUs $\{b_1, b_2, ..., b_{\#\mathrm{CUs}}\}$
        \State $\beta=\mathrm{memory\_bandwidth} / \mathrm{element\_size}$
        \State $\gamma=\mathrm{output\_buffer\_size}$
        \State $\#\mathrm{runs}=\#\mathrm{batches}*(M*pk)/(\gamma*\#\mathrm{CUs})$
        \For{\#runs}
            \For{$\gamma/\beta$}
                \For{$b$ in $\{b_1, b_2, ..., \#\mathrm{CUs}\}$}
                    \State $\mathcal{M} \leftarrow \beta$ elements from $b$
                \EndFor
            \EndFor
        \EndFor
    \end{algorithmic}
\end{algorithm}

Once the tile buffers are loaded, the data within them are streamed into the computation unit. The results are then transposed and streamed to the off-chip memory via a ping-pong buffer in the data mover. The computation unit is a systolic array composed of finite field processing elements that carry out multiplication and accumulation within a finite field. This systolic array enables the pipelining and parallelization of tile multiplication. Consequently, the level of parallelism increases with the tile size ($t_m\times t_n$), though this also results in higher resource consumption.

In addition to computing the elements of a tile in parallel, the generation of each batch is independent and can also be parallelized. This batch-level parallelism is exploited by replicating multiple BATS CUs on the FPGA. Ideally, each BATS CU should operate fully independently to achieve maximum parallelism, meaning that there should be no communication or contention among the CUs. However, I/O operations often become a bottleneck, leading to potential stalling. In the following discussion, we will explore the design of the input and output mechanisms and examine strategies to minimize contention.

\subsubsection{Input Port Contention and Delay}
The Xilinx U50 board is equipped with an 8 GB HBM2 stack, organized into 32 pseudo channels (PCs) (DDR memory banks), as illustrated in \cref{fig:block}. 
A BATS CU reads data from the HBM through three components: an HBM pseudo channel, a crossbar switch and an AXI adapter. Input contention among CUs and delay occur in these tree components.

In our proposed design, each BATS CU has its own AXI adapter to access the HBM PCs for input data. It is important to note that while bundling the ports of multiple BATS CUs into a single AXI adapter could be considered if resource constraints are a concern, this approach would likely lead to significant contention and queueing delay as the AXI adapter is not a hardened unit.

Four pseudo channels (PCs) share a crossbar switch within the HBM, allowing an AXI channel to access any memory banks either through internal connections of the crossbar switch or through lateral connections, as depicted in \cref{fig:block}(c). The most efficient access occurs within a single crossbar switch, known as aligned access. Accessing memory banks across different crossbar switches results in longer delays compared to within a single switch. According to AMD's official documentation~\cite{crossbar}, the maximum latency for aligned access is 90 clock cycles, whereas the minimum latency for unaligned access is 110 clock cycles, with the potential for even higher latency depending on data traffic across the switches.

Therefore, the most efficient HBM configuration involves creating individual copies of the input data across different PCs to allow each BATS CU to operate on its own data without contention. For instance, as illustrated in \cref{fig:block}(c), with four BATS CUs, we store four copies of the input data in PCs 0, 1, 2, and 3, respectively. Each BATS CU then accesses its data through dedicated AXI adapters: M0-PC0, M1-PC1, M2-PC2, and M3-PC3. This configuration minimizes contention and maximizes access efficiency.
However, this approach increases HBM consumption and data copy time. A compromise to better balance overall performance with HBM usage is to share a data copy among PCs within the same crossbar switch. For example, we can use PC0 to store the data, and each BATS CU would then access the data through M0-PC0, M1-PC0, M2-PC0, and M3-PC0. This configuration reduces the number of PCs and input data copies by a factor of four, while introducing some contention among CUs. Nevertheless, the contention from sharing PCs within the same crossbar switch is minimal compared with sharing AXI adapters, as the HBM operates at a significantly higher clock speed than the AXI ports~\cite{crossbar}.

\subsubsection{Output Port Contention}
Ideally, each BATS CU should have an individual output port and transfer generated data to HBM independently without stalling. However, the number of memory ports in HBM is limited, and an output port must be shared across multiple CUs. Therefore, the contention of accessing the output port is inevitable, and CUs will be stalled if the output logic is not designed carefully.

Unlike the input, where the entire system may stall if data is unavailable, output contention can be mitigated through buffering results from the compute logic. Specifically, each CU uses two \(t_m \times M\) ping-pong buffers, as detailed in \cref{alg:cu-ping-pong}. Tiled results of size \(t_m \times t_n\) from the systolic array are transposed and stored in one of the buffers. When one buffer becomes full, its data is streamed to the HBM at full memory bandwidth, while the other buffer simultaneously begins receiving data from the systolic array.

Since the generated results are buffered on-chip and HBM bandwidth can readily be fully utilized, an output port can be shared among multiple CUs, as outlined in \cref{alg:data-out}. However, to prevent stalling, especially as the number of CUs or tile size increases, additional output ports may be necessary. Further details on the required number of output ports will be provided in \cref{sec:imp}.

\subsection{Load Balance Scheduler}

\begin{figure}
    \centering
    \includegraphics[width=0.95\linewidth]{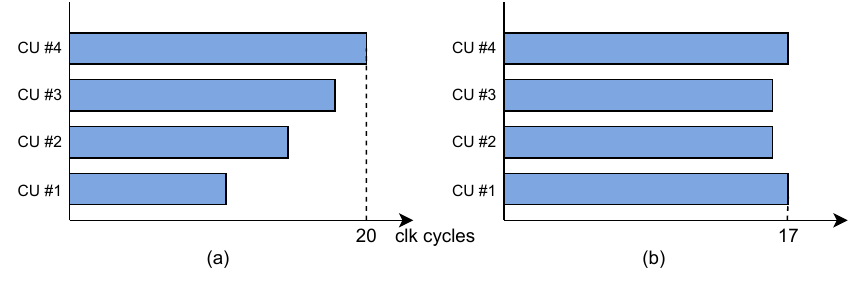}
    \caption[Load balance comparision]{\textbf{Scheduling comparison.} Runtime comparison of 4 CUs is shown. The clock cycles shown are for discussion purposes. (a) Sequential scheduling. (b) Load balance scheduling. }
    \label{fig:load-balance}
\end{figure}

The BATS CUs need to be scheduled according to the number of batches required. 
A straightforward approach is to schedule them sequentially, where the CUs process the rows of the base graph one by one.
For example, consider the rows degrees of one of the base graphs in \cite{cs-bats}: 
\begin{equation*}
    \{11, 12, 14, 14, 16, 19, 20, 27\}.
\end{equation*}
In the above, the row degrees are listed in ascending order. 
If four BATS CUs are scheduled sequentially to generate 16 batches according to this base graph, the row degrees assigned for each CU will be 
\begin{align*}
    &\textnormal{CU \#1:}\ 11,\ 16, \ 11, \ 16 ;\\
    &\textnormal{CU \#2:}\ 12,\ 19, \ 12, \ 19 ;\\
    &\textnormal{CU \#3:}\ 14,\ 20, \ 14, \ 20 ;\\
    &\textnormal{CU \#4:}\ 14,\ 27, \ 14, \ 27,
\end{align*}
As a result,
CU~\#1 will process two rows with degrees of 11 and 16 twice, CU~\#2 will process two rows with degrees of 12 and 19 twice, CU~\#3 will process two rows with degrees of 14 and 20 twice, and CU~\#4 will process two rows with degrees of 14 and 27 twice. 
As we can see, sequential scheduling results in uneven runtimes across the CUs, leading to a longer overall runtime, as illustrated in \cref{fig:load-balance}(a).
Notice that in this example, the number of CUs divides the number of rows in the base graph for discussion purposes. This condition, however, is not necessary in practice.

\begin{algorithm}
    \caption{Load balance scheduling}\label{alg:load-balance}
    \begin{algorithmic}[1]
        \State \textbf{Input} Base graph $\mathcal{G}=\{g_1, ..., g_{m}\}$; BATS CUs $\{c_1, c_2, ..., c_{\#\mathrm{CUs}}\}$; number of batches $N$
        \For{$i$ in $\{1, 2, ..., \lceil{N / \#\mathrm{CUs}}\rceil\}$}
            \For{$j$ in $\{1, 2, ..., \#\mathrm{CUs}\}$}
                \State batch\_id  $\leftarrow(i-1)*\#\mathrm{CUs} + j$
                \If{$\mathrm{batch\_id} \leq N$}
                    \State $g\leftarrow g_{(\mathrm{batch\_id} \bmod{m})} \ggg \lfloor{\mathrm{batch\_id} / m}\rfloor$
                    \If{$(\mathrm{batch\_id} \bmod{m}) = 1$}
                        \State $j \leftarrow \#\mathrm{CUs} - j + 1$
                    \EndIf
                    \State use $c_{j}$ to run $g$
                \EndIf
            \EndFor
        \EndFor
    \end{algorithmic}
\end{algorithm}

One solution to address the issue of uneven runtimes is to interleave the rows of the base graph. However, this approach would require altering the original coding scheme. To avoid unnecessary modifications to the original algorithm, we propose the approach outlined in \cref{alg:load-balance}, which schedules the BATS CUs by pairing a high degree with a low degree in order to balance the workload.
Specifically, after generating every \(m\) batches (or one layer of the CS-BATS), we reverse the scheduling order of the CUs. This means that the CUs responsible for generating batches with smaller degrees in the previous layer will be scheduled to handle batches with larger degrees in the new layer, and vice versa. This method balances the workload across all BATS CUs without altering the structure of the original CS-BATS, resulting in a shorter total runtime and higher hardware utilization, as shown in \cref{fig:load-balance}(b).

\section{Bounded-value Generators}
The throughput and resource consumption of a BATS accelerator on FPGA are primarily influenced by two critical factors. The first factor is how data flow through the accelerator, encompassing aspects such as data I/O, buffering, pipelining, and the overlapping of different processing stages. The second factor is the design of the finite field multiplier.

In the compute unit, we parallelize \(t_m \times t_n\) finite field multiply-accumulate (MAC) operations within a single clock cycle, which requires \(t_m \times t_n\) MAC units in each BATS CU. Assuming the use of a binary extension field, \(GF(2^n)\), the accumulation in the finite field simplifies to an exclusive-or (XOR) operation. Consequently, each MAC unit is primarily composed of a finite field multiplier and an XOR operator.

While the previous sections focused on the design of individual modules and the overall data flow, this section will focus on optimizing the finite field multiplier to enhance performance and efficiency.

\subsection{Finite Field Multiplication on FPGA}
When designing a coding or cryptographic system, the theoretical performance is significantly influenced by the field size used in the basic arithmetic operations. In other words, if we consider only the theoretical aspects and ignore the efficiency of performing field operations on a software or hardware platform, the algorithm's performance will be identical as long as the chosen fields are isomorphic.
However, the choice of field can significantly impact the system's performance, particularly on hardware, even when two fields are isomorphic or of similar size. For instance, extension fields with smaller characteristics, especially binary extension fields, are generally more advantageous than a prime field of comparable size, such as \(GF(p)\), where \(p \approx 2^n\). It is because addition in a binary extension field can be efficiently implemented with a multi-input XOR gate that has a short delay, whereas addition in a prime field is more complex due to the longer carry propagation required~\cite{gf-survey}. Consequently, a \(GF(p)\) multiplier is inherently more complex than a \(GF(2^n)\) multiplier. Furthermore, elements of \(GF(2^n)\) can be represented as an \(n\)-dimensional vector space over \(GF(2)\), which naturally aligns with the conventional binary data representation. Therefore, in this work, we focus on implementing arithmetic operations in \(GF(2^n)\).      

\subsubsection{Two approaches}
There are two primary approaches to implementing a finite field multiplier: the look-up table approach and the direct computation approach. In the following, we analyze the complexity of each approach. 

\subsubsection{Complexity of look-up table}
The look-up table method optimizes multiplication by converting the operation into two logarithmic look-ups, one anti-logarithmic look-up, and one XOR operation, necessitating two look-up tables of size \(2^n\) elements each. This approach is particularly advantageous in software implementations running on general-purpose processors, where the available runtime memory is typically much larger than the on-chip memory of an FPGA. Additionally, with the availability of Single-Instruction-Multiple-Data (SIMD) instructions in some CPU architectures, multiple MAC operations can be executed in parallel.

However, the two look-up tables, each with $2^n$ elements, must be replicated for each MAC unit to enable parallel processing.  
This replication can be resource-intensive and inefficient for FPGA, which has limited on-chip storage compared with CPU. 

The complexity of implementing on-the-fly multiplication of two finite field elements depends on the choice of basis used to represent the field elements. Common bases include polynomial basis, normal basis, triangular basis, and redundant basis~\cite{gf-rb}, with the polynomial basis generally offering the lowest space-time complexity~\cite{gf-survey}. 

Multiplication using a polynomial basis typically involves two stages: multiplication and reduction. First, polynomial multiplication is performed, followed by a modulo operation with a degree-\(n\) irreducible polynomial that defines the field. 

In this work, we will focus on optimizing the direct computation approach for finite field multiplication using the polynomial basis.

\subsubsection{Complexity of direct computation}
\begin{algorithm}
    \caption{Multiplication in $GF(2^n)$}\label{alg:gf-mul}
    \begin{algorithmic}[1]
        \State \textbf{Input} $a$ and $b\in GF(2^n)$ with irreducible\_polynomial
        \State reg $\leftarrow 0$
        \State ta $\leftarrow a$
        \For{$i$ in $\{1, 2, ..., n\}$}
            \If{the $i$-th bit of $b$ is 1}
                \State reg $\leftarrow \mathrm{reg} \oplus \mathrm{ta}$ 
            \EndIf
            \If{the $n$-th bit of ta is 1}
                \State ta $\leftarrow \mathrm{irreducible\_polynomial} \oplus (\mathrm{ta} \ll 1)$ 
            \Else
                \State ta $\leftarrow \mathrm{ta} \ll 1$ 
            \EndIf
        \EndFor
        \State \Return reg
    \end{algorithmic}
\end{algorithm}

\begin{figure}
    \centering
    \includegraphics[width=1\linewidth]{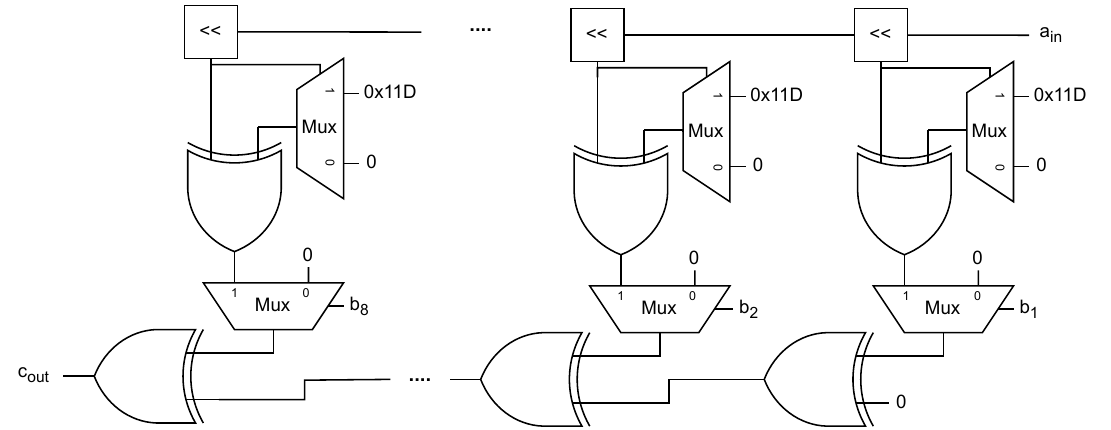}
    \caption[Multiplier chain]{\textbf{Multiplier chain of a bit-parallel $GF(2^8)$ multiplier.} Three out of eight partial multipliers are shown. Two operands, $a$ and $b$, will be multiplied, where $a$ will be input through $a_{in}$. Operand $b$ will be used to generate control signals for the multiplexers. XOR gates shown perform exclusive-or operation between two 8-bit numbers. Irreducible polynomial $x^8+x^4+x^3+x^2+1$ is used in this example.}
    \label{fig:gf_mul}
\end{figure}

A common implementation of finite field multiplication is based on the ancient Egyptian multiplication algorithm. In this algorithm, one number is iteratively doubled while the other is iteratively halved, with the doubled number being accumulated if the halved number is odd~\cite{algo-mul}. This approach can be adapted to \(GF(2^n)\), as described in \cref{alg:gf-mul}.

Specifically, given two operands \(a\) and \(b\) from \(GF(2^n)\), represented by polynomials \(f(x)\) and \(g(x)\) respectively, the algorithm proceeds as follows: First, initialize an accumulation register to zero. Then, For $n$ iterations, we do the following steps.
Step 1) Check if \(b\) is divisible by \(x\). If not divisible, the current value of \(a\) is added to the accumulation register. Otherwise, proceed to the next step.
Step 2) 
\(a\) is iteratively multiplied by \(x\) (effectively doubling it in the field) and then reduced modulo the irreducible polynomial, and \(b\) is iteratively divided by \(x\) (halving it). 

This algorithm can be implemented as a bit-parallel multiplier by unrolling the loop in \cref{alg:gf-mul}, effectively creating a chain of \(n\) partial multipliers. These partial multipliers perform accumulation, shifting, and reduction on the operand \(a\) based on the operand \(b\) during different iterations, as illustrated in \cref{fig:gf_mul}. By unrolling the loop, the \(n\)-bit finite field multiplication can be completed in a single clock cycle.

If we measure the space complexity by the number of 1-bit XOR gates, \cref{alg:gf-mul} requires \(2n^2\) XOR gates, resulting in a quadratic space complexity, \(O(n^2)\). Similarly, the propagation delay can be measured by the delay of a single 1-bit XOR gate. Since the two \(n\)-bit XOR operations in \cref{alg:gf-mul} can be executed in parallel, and the 1-bit XOR operations within an \(n\)-bit XOR can also be performed in parallel, the time delay of each partial multiplier is \(O(1)\). This results in a total propagation delay complexity of \(O(n)\).

Most related work focuses on optimizing either the polynomial multiplication stage or the reduction stage of finite field multiplication~\cite{gf-survey, gf-opt, gf-opt2}. In this work, we adopt a straightforward yet effective approach to significantly reduce the space-time complexity of a finite field multiplier by an order of magnitude. Although this approach is specifically designed for the BATS code, it has the potential to be generalized to other block codes that employ similar operations.

\subsection{Bounded-value Coding Coefficients}
For a binary extension field with a polynomial basis, $GF(2^n)$, we define its bounded-value subset of length $s$ as follows. 
\begin{definition}
(Bounded-value subset)
Let $L(2^s)$ be the subset of $GF(2^n)$ that contains all elements in $GF(2^n)$ such that its binary representation is in the range $0, 1, ..., 2^s-1$, where $s\leq n$.
\label{def:subfield}
\end{definition}
With the binary representation, this subset contains all elements from $GF(2^n)$, with the most significant $n-s$ bits being zeros. Thus, elements from this subset can be represented by $s$ bits.
A generator matrix with values taken only from this bounded-value subset is called a bounded-value (BV) generator.

\begin{algorithm}
    \caption{Multiplication with BV elements}\label{alg:gf-mul-subfield}
    \begin{algorithmic}[1]
        \State \textbf{Input} $a\in GF(2^n)$ with irreducible\_polynomial, $b\in L(2^s)$
        \State reg $\leftarrow 0$
        \State ta $\leftarrow a$
        \For{$i$ in $\{1, 2, ..., s\}$}
            \If{the $i$-th bit of $b$ is 1}
                \State reg $\leftarrow \mathrm{reg} \oplus \mathrm{ta}$ 
            \EndIf
            \If{the $n$-th bit of ta is 1}
                \State ta $\leftarrow \mathrm{irreducible\_polynomial} \oplus (\mathrm{ta} \ll 1)$ 
            \Else
                \State ta $\leftarrow \mathrm{ta} \ll 1$ 
            \EndIf
        \EndFor
        \State \textbf{Return} reg
    \end{algorithmic}
\end{algorithm}

In \cref{def:subfield}, we introduce the concept of a bounded-value (BV) subset. 
Recall that a BATS code is defined by the linear system \(\mathbf{X}_i = \mathbf{B}_i \mathbf{G}_i\), where \(\mathbf{G}_i\) is the generator matrix that specifies the linear combinations applied to the selected input packets. 
In the standard construction of a BATS code, the entries of all \(\mathbf{X}_i\), \(\mathbf{B}_i\), and \(\mathbf{G}_i\) are taken from \(GF(2^n)\). 
Here, we propose to restrict the entries of the matrix \(\mathbf{G}_i\) to a BV subset \(L(2^s)\). 

Using a BV generator in the BATS code construction has several advantages:
\begin{enumerate}
    \item Reduced Logic Complexity: In the original multiplication algorithm \cref{alg:gf-mul}, multiplying two elements \(a\) and \(b\) from \(GF(2^n)\) involves left-shifting and accumulating \(a\) according to each bit of \(b\). By restricting \(b\) to the BV subset \(L(2^s)\), only \(s\) bits need to be checked instead of \(n\) in the multiplication loop (\cref{alg:gf-mul-subfield}). This reduction in bit operations decreases the number of partial multipliers needed in the multiplier chain, leading to significant logic resource savings on the FPGA.
    \item Shorter Propagation Delay: With a shorter multiplier chain, the propagation delay is reduced. This results in a potentially shorter critical path delay, which improves the overall timing and performance of the design during synthesis. This is particularly beneficial for high-speed implementations.

    \item Efficient Resource Usage: The BV generator, requiring fewer bits, can be stored in a smaller on-chip ROM, saving hardware resources. This efficiency enhances the scalability of the design, allowing for more complex or larger implementations without exhausting available resources.
\end{enumerate} 
These advantages make the use of BV generators a compelling choice for optimizing BATS code implementations on FPGA platforms.

The concept of BV generators draws inspiration from the research on efficient large language model (LLM) deployment, where techniques like parameter quantization~\cite{quant} and knowledge distillation~\cite{dist} are employed to reduce the size of a trained LLM, enabling efficient deployment across various computational platforms. These methods essentially involve using a reduced subset of parameters or neural network layers to replace the original large model, with a focus on minimizing the impact on model performance.

Similarly, the design of the BV generators in the BATS code construction follows this philosophy. By restricting the generator matrix \(\mathbf{G}_i \) to a BV subset \(L(2^s)\), the size and complexity of the original code structure are reduced while maintaining an acceptable or essentially unaffected performance level (to be discussed in the next subsection). 
This approach allows for more efficient use of hardware resources, such as logic elements and on-chip memory, making the implementation more scalable and better suited for deployment on FPGA platforms.

In the following discussion, we will delve into the design considerations for BV generators, exploring how the choice of BV subset size \(s\) impacts the performance of the BATS code. The goal is to find a good balance between resource efficiency and code performance, much like the trade-offs considered in LLM deployment techniques. 
In fact, we will show that the BV generator can have almost no impact on the coding performance. 

\section{Impact on Code Performance}
In this section, we will first derive an analytical expression for the probability of rank deficiency, a key factor influencing the performance of BATS codes, in relation to the BV subset size.  
Subsequently, simulation results will be presented to identify the optimal BV subset size for practical applications, corroborating the theoretical analysis.  
Lastly, we will conduct a detailed complexity analysis and compare it with the conventional method, focusing on the optimal BV subset size.

\subsection{Theoretical analysis}
Recall that the generated batches will be transmitted through a multihop network\footnote{For simplicity, we consider a multihop line network.} with packet loss. During the transmission across each hop, packets within a batch may be lost. The received packets at each hop will be linearly recombined or ``recoded'' into a new batch of size $M$. At the destination node, each received batch is treated as a system of linear equations, which can be solved when the system's rank matches its degree. To mitigate decoding complexity, belief propagation decoding is typically employed; this method iteratively solves each batch locally and substitutes the solved variables into other batches.
In general, the end-to-end system for the $i$-th batch can be described by 
\begin{equation}
    Y_i = B_iG_iH_i, \label{eq:bats-sys}
\end{equation}
where $H_i$ is used to describe the packet loss and the recoding process. For example, a packet loss can be represented by a zero row in $H_i$. 

In the following discussion, the batch index $i$ will be omitted where there is no ambiguity. In \cref{eq:bats-sys}, $Y$ represents the received packet, $G$ is known, and $H$ can be obtained by transmitting an identity matrix along with the batch from the source node.
Let $G'=GH$. To solve for $B$, it is crucial that $\mathrm{rank}(G')$ equals $\mathrm{degree}(B)$. Note that the rank of $G\in \mathbb{F}_q^{dg\times M}$ is determined by the construction algorithm. 
By post-multiplying $G$ by $H$ to obtain $G'$, the rank of $G$ may decrease due to two factors. The first factor is packet loss, which is governed by the channel conditions. The second factor is rank deficiency, arising from the finite number of elements in the Galois field. Although we can ensure that $G$ has full rank at the time of generation, rank deficiency may still occur if a column of $G$ is removed due to packet loss during transmission. 
It is evident that restricting $G$ to a BV subset with fewer elements increases the likelihood of rank deficiency. 
The questions are how significant this impact is and how it is related to the size of the BV subset.

\subsubsection{Full rank probability over a subset}

Consider a finite field with $q$ elements. The probability of a randomly generated $m$ by $r$ matrix in this field to be full rank is given by 
\begin{equation}
    \zeta^m_r(q) \triangleq (1-q^{-m})(1-q^{-m+1})\cdots(1-q^{-m+r-1}),\label{eq:full-rank-prob}
\end{equation}
where $0<r\leq m$~\cite{bats-tit}. 
Consider a subset of $GF(q)$ with $u\leq q$ elements including $0$. The probability of a randomly generated matrix in this subset to be full rank has a lower bound of a similar form, namely, 
\begin{equation}
    \zeta^m_r(u) \geq (1-u^{-m})(1-u^{-m+1})\cdots(1-u^{-m+r-1}).\label{eq:full-rank-prob-u}
\end{equation}
It is obvious that (\ref{eq:full-rank-prob}) can be derived by counting the number of non-zero vectors and chosen vectors of size $m$ in the finite field of $q$ elements.
To prove (\ref{eq:full-rank-prob-u}), we notice the following. 
Let $U$ be the subset of $GF(q)$ of size $u$ including $0$.
To form a full-rank matrix using elements in the subset of size $u$, we also need to select $r$ linearly independent vectors in sequence.
The first column vector must be non-zero. There are $u^m - 1$ non-zero vectors in $U^m$. Since there are $u^m$ total vectors in $U^m$, the probability of choosing a valid first column vector is:
\begin{equation}
   P_1 = \frac{u^m - 1}{u^m} = 1 - u^{-m}.
\end{equation}

The second column vector must be linearly independent of the first. Let $\alpha$ denote the number of vectors in $U^m$ that lie in the span of the first column vector. 
Notice that we have $\alpha \leq u$. 
To justify the bound $\alpha \leq u$:
\begin{itemize}
   \item For any non-zero vector $v \in U^m$, consider the set of all scalar multiples $\{\gamma \cdot v \mid \gamma \in GF(q)\}$. A scalar multiple $\gamma \cdot v \in U^m$ if and only if all components of $\gamma \cdot v$ belong to $U$.
   \item For a given component $v_i \in U$, the condition $\gamma v_i \in U$ depends on whether $v_i = 0$ or $v_i \neq 0$:
       \begin{itemize}
           \item If $v_i = 0$, then $\gamma v_i = 0 \in U$ for any $\gamma \in GF(q)$, and no restriction is imposed on $\gamma$ by this component.
           \item If $v_i \neq 0$, the condition $\gamma v_i \in U$ restricts $\gamma$ to belong to the set $U \cdot v_i^{-1}$ (where $v_i^{-1}$ is the multiplicative inverse of $v_i$ in $GF(q)$). Since $|U| = u$, there are at most $u$ valid choices for $\gamma$.
       \end{itemize}
   \item Extending this to the entire vector $v$, the intersection of restrictions across all non-zero components ensures that the number of valid scalars $\gamma$ for which $\gamma \cdot v \in U^m$ is at most $u$, regardless of the structure of $U$ (whether it is a subfield or not).
\end{itemize}
Therefore, the number of vectors in the span of the first column vector that belong to $U^m$ is at most $u$. Thus, $\alpha \leq u$. The number of valid choices for the second column vector is at least $u^m - \alpha \geq u^m - u$, and the probability of choosing a valid second column vector is:
\begin{equation}
   P_2 \geq \frac{u^m - u}{u^m} = 1 - u^{-m+1}.
\end{equation}

Repeating this process for each subsequent column, the $j$-th column must be linearly independent of the span of the previous $j-1$ columns. In the worst case, the number of vectors in $U^m$ that lie in the span of the first $j-1$ columns is bounded by $u^{j-1}$. Thus, the number of valid choices for the $j$-th column is at least $u^m - u^{j-1}$. The probability of choosing a valid $j$-th column vector is at least:
\begin{equation}
   P_j \geq \frac{u^m - u^{j-1}}{u^m} = 1 - u^{-m+j-1}.
\end{equation}
The matrix will be full rank if all $r$ column vectors are chosen correctly. By the multiplication rule for probabilities and the independence of choices, the overall probability is:
\begin{equation}
\begin{aligned}
    \zeta^m_r(u) &\geq P_1 \cdot P_2 \cdot \ldots \cdot P_r \\
                 &\geq (1 - u^{-m})(1 - u^{-m+1})\cdots(1 - u^{-m+r-1}).
\end{aligned}
\end{equation}
This completes the proof for (\ref{eq:full-rank-prob-u}).    

\subsubsection{Rank deficiency probability after deletion}
Consider two matrices, $G$ and $G'$, where $G$ is a randomly generated matrix in a subset of $GF(q)$ with $u$ elements, and $G'$ is obtained by randomly removing one column of $G$. 
Let us define two events,
\begin{equation*}
    A: G \in \mathbb{F}^{\mathrm{dg}\cdot M}_q \quad\textnormal{is full rank},
\end{equation*}
\begin{equation*}
    B: G' \in \mathbb{F}^{\mathrm{dg}\cdot (M-1)}_q \quad\textnormal{is full rank}.
\end{equation*}
To investigate the impact of the field size on rank deficiency probability for $G$, we first derive the full rank probability when one column of $G$ is deleted, namely $P(B|A)$.

Firstly, we notice that when $\mathrm{dg} \geq M$, $P(B|A)=1$, because all vectors in the column space of $G$ are linearly independent. Therefore, we consider $\mathrm{dg}<M$. By Bayes' theorem, we have 
\begin{equation}
    P(B|A) = \frac{P(B)P(A|B)}{P(A)}.
\end{equation}
Notice that when $\mathrm{dg}<M$, we have $P(A|B)=1$ for the following reason.
If $B$ is given, we have $rank(G')=\mathrm{dg}$, because $\mathrm{dg}<M$ implies $\mathrm{dg}\leq M-1$. Since $\mathrm{dg}\geq rank(G)\geq rank(G')=\mathrm{dg}$, we obtain $rank(G)=\mathrm{dg}$, which means that $G$ is full rank whenever $G'$ is full rank, so that $P(A|B)=1$. 
Thus, we can write
\begin{equation}
    P(B|A) = \frac{P(B)}{P(A)} = \frac{\zeta^{M-1}_{\mathrm{dg}}(u)}{\zeta^{M}_{\mathrm{dg}}(u)} \geq \frac{1-u^{-M+\mathrm{dg}}}{1-u^{-M}}. \label{eq:pba}
\end{equation}
Likewise, we can define $G''\in \mathbb{F}^{\mathrm{dg}\cdot (M-2)}_q$ and the event 
\begin{equation*}
    C: G'' \in \mathbb{F}^{\mathrm{dg}\cdot (M-2)}_q \quad\textnormal{is full rank}.
\end{equation*}
With a similar analysis, for $\mathrm{dg}<M-1$, we can obtain
\begin{equation}
    P(C|A) \geq \frac{(1-u^{-M+\mathrm{dg}})(1-u^{-M+\mathrm{dg}+1})}{(1-u^{-M})(1-u^{-M+1})}. \label{eq:pca}
\end{equation}

\begin{figure}
    \centering
    \includegraphics[width=0.7\linewidth]{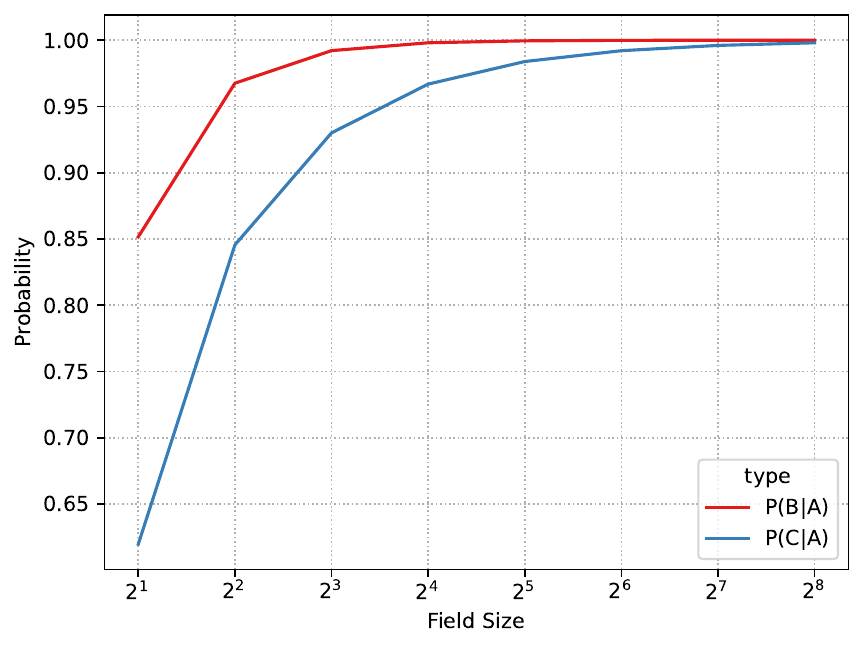}
    \caption{Visualization of the lower bound for the full-rank probability, $P(B|A)$ and $P(C|A)$, against the subset sizes $u$ after columns are deleted from $G$. The average results for generator matrices with degrees $\{11, 12, 14, 14\}$ and $M=16$ are presented.}
    \label{fig:prob}
\end{figure}

In \cref{fig:prob}, we plot the lower bounds in (\ref{eq:pba}) and (\ref{eq:pca}) against various subset sizes, where $M$ is set to 16 and the base graph degrees from \cite{cs-bats} are used, namely the generators with degrees $\{11, 12, 14, 14\}$. In general, the lower bound for the probability of achieving full rank (\ie, 1 minus the rank deficiency probability) increases with the subset size. Although a small subset size, such as $2^1$, results in a rather small lower bound for the full-rank probability (only 62\% chance of being full-rank after two packets are lost), some relatively small subset sizes, such as $2^4$, offer a higher lower bound for the full-rank probability comparable to that of using the entire finite field.
This suggests that we can use elements from a relatively small subset rather than the entire finite field to construct the generator matrix. 

\subsection{Simulations}

To illustrate the impact of the BV subset size more directly, we conduct simulations of the CS-BATS code using BV generators of varying sizes.

The simulation setup follows the methodology outlined in \cite{cs-bats}, using $GF(2^8)$ with a packet number $K=256$, packet size $pk=256$, and batch size $M=16$. The base graph is generated using the same procedure and degrees, namely \{11, 12, 14, 14, 19, 20, 27\}. A 10\% i.i.d.\ packet loss is applied at each hop during the simulation. Two types of experiments are conducted, similar to those in \cite{cs-bats}. 

In Experiment~1, we examine the effect of the number of hops on the decoding rate, with the number of batches fixed at 20.

In Experiment~2, we fix the number of hops at 10 and vary the number of batches. Each simulation is repeated 500 times, and the average results are reported.

\begin{figure}
    \centering
    \subfloat[Inactivation Decoding]{\includegraphics[width=0.5\linewidth]{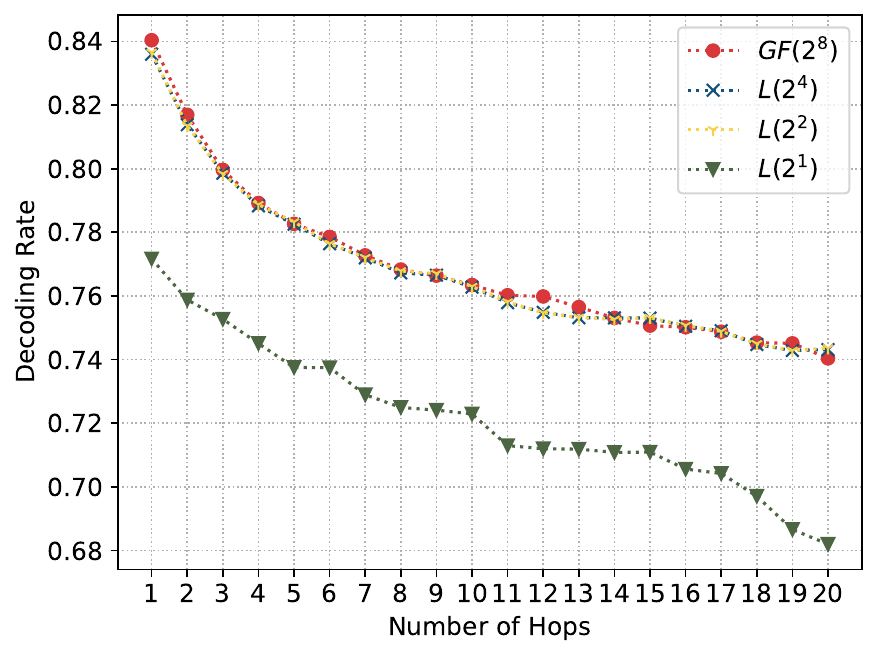}\label{fig:decoding_rates1a_fpt}}
    \subfloat[BP Decoding]{\includegraphics[width=0.5\linewidth]{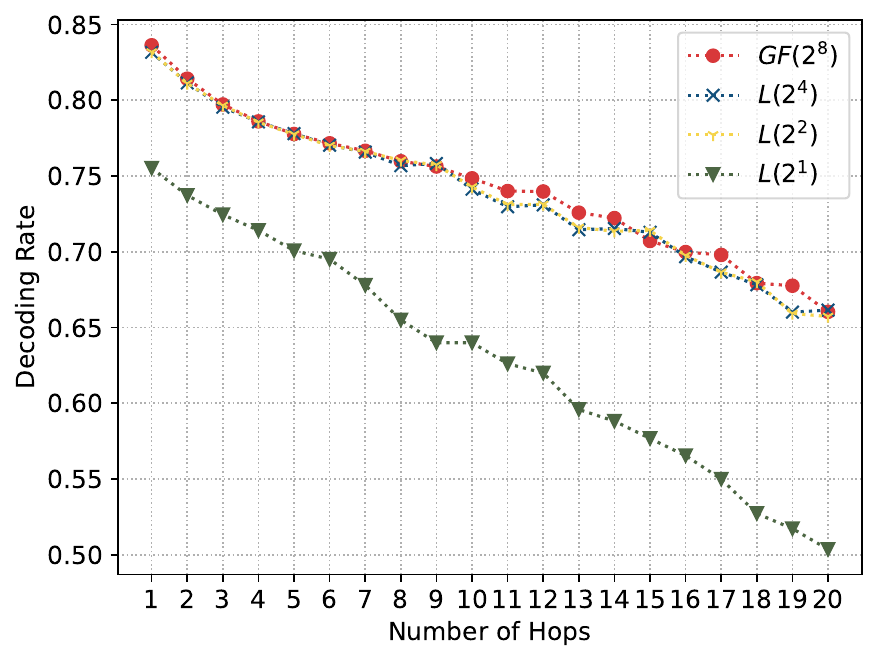}\label{fig:decoding_rates1b_fpt}}
    
    \caption[Experiment 1]{\textbf{Experiment 1.} Decoding rates vs the number of hops using (a) inactivation decoding and (b) BP decoding. The average results from 500 trials are presented. $L(2^4), L(2^2), L(2^1)$: CS-BATS using BV generators with various BV subset sizes. $GF(2^8)$: CS-BATS with regular generators. }
    \label{fig:decoding_rates1_fpt}
\end{figure}

\begin{figure}
    \centering
    \subfloat[Inactivation Decoding]{\includegraphics[width=0.5\linewidth]{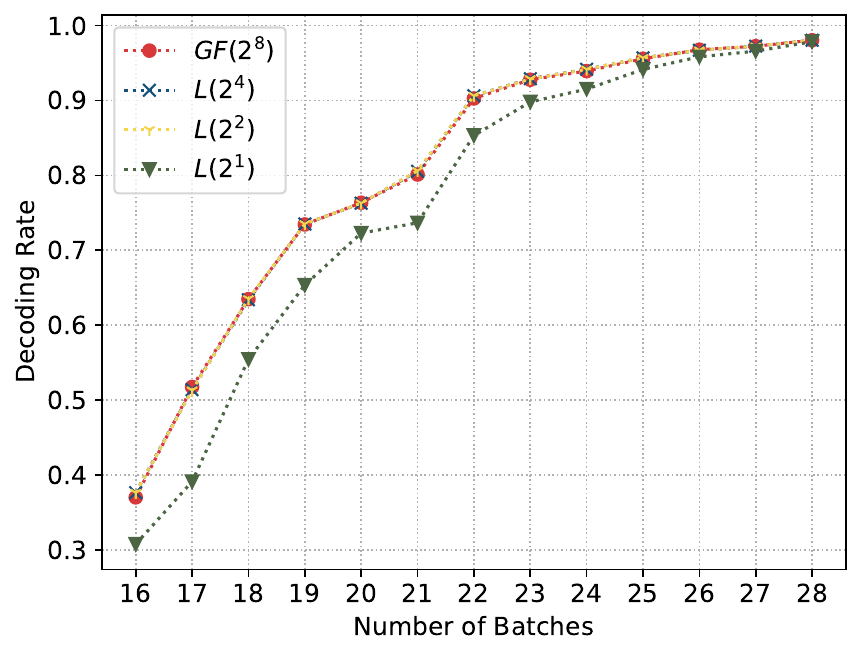}\label{fig:decoding_rates2a_ftp}}
    \subfloat[BP Decoding]{\includegraphics[width=0.5\linewidth]{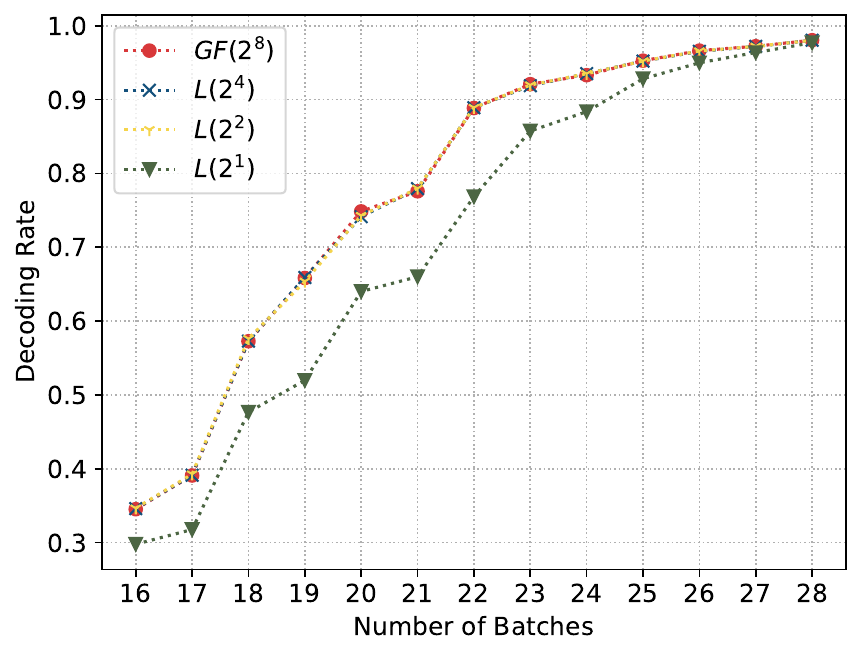}\label{fig:decoding_rates2b_ftp}}
    \caption[Experiment 2]{\textbf{Experiment 2.} Decoding rates vs the number of batches using (a) inactivation decoding and (b) BP decoding. The average results from 500 trials are presented. $L(2^4), L(2^2), L(2^1)$: CS-BATS using BV generators with various BV subset sizes. $GF(2^8)$: CS-BATS with regular generators.}
    \label{fig:decoding_rates2_fpt}
\end{figure}

The simulation results are presented in \cref{fig:decoding_rates1_fpt} and \cref{fig:decoding_rates2_fpt}, comparing the performance of using BV generators with $L(2^4)$, $L(2^2)$, and $L(2^1)$ against not using BV generators. 

Two decoding strategies, also as outlined in \cite{cs-bats}, are simulated in both experiments. In the figures, the red lines represent the conventional CS-BATS code with the generator matrix in $GF(2^8)$, while the other lines represent the CS-BATS code using BV generators. 

It is evident that, except for $L(2^1)$, the use of BV generators yields results nearly identical to the results using the entire finite field. When BV generators in $L(2^1)$ are employed, performance deteriorates significantly across all experiments. This observation aligns with our previous analysis, which shows a sharp decline in the full-rank probability after two hops when using $L(2^1)$, as depicted in \cref{fig:prob}.

\subsection{Savings analysis}
According to our simulations, the smallest BV subset size that can be used without compromising performance under the discussed setup is $L(2^2)$. 

Recall that each BATS CU requires $t_m \times t_n$ MAC units to process the elements in a tile in parallel. The conventional method necessitates $2n^2$ 1-bit XOR gates for each multiplier. Consequently, a MAC unit requires $2n^2 + n$ 1-bit XOR gates, accounting for the $n$-bit accumulation needed after multiplication. 

For a tile size of $8 \times 8$ and $n=8$, the total number of 1-bit XOR gates required for the conventional method amounts to 8704.
On the other hand, using a BV generator in $L(2^s)$ reduces the number of 1-bit XOR gates required per MAC unit to $2ns + n$. For $s=2$, the total number of 1-bit XOR gates needed is 2560, which represents a 3.4-fold reduction in logic resources compared with the conventional method. 

Even with a more robust design using $s=4$, the logic resource requirement is approximately half of that needed for the conventional method. Additionally, the propagation delay for each MAC unit is also reduced when BV generators are employed.

Recall that the generator matrices are pre-generated and stored in the on-chip ROM. For the base graph with 7 rows, as described, the conventional method requires $\sum_i dg_i \times M = 1872$ bytes of ROM. In contrast, when using a BV generator in $L(2^2)$, only 468 bytes of ROM are needed to store all the generator matrices on the FPGA.

\section{Implementation Results}
\label{sec:imp}
In this section, we proceed with the implementation of the BATS code accelerator as previously outlined. Specifically, we conduct a comparative analysis of the implementations with and without BV generators, highlighting the impact of the BV subset size on the overall resource utilization.  
Additionally, we examine various factors affecting the scalability of the implementation, including port bundling, HBM PC configurations, and load balancing strategies.

All implementations were carried out using the AMD Vitis Unified IDE 2023.1, targeting the XCU50-FSVH2104-2-E device. The Vitis flow was utilized throughout the implementation process. Specifically, the design undergoes simulation and synthesis using high-level synthesis (HLS). Once the C/RTL co-simulation and synthesis results satisfy the design requirements, the implementation phase is initiated.  
A system project is then created, incorporating a customized OpenCL-based application to serve as the host-side driver for executing the design on board. The design is deployed on an Alveo U50 board, connected via a PCIe interface to a ThinkStation-P350 computer equipped with an Intel I9-10900 processor.  
The HBM is configured in accordance with our experimental setup within the hardware link configuration. Kernel runtime profiling and memory profiling are both configured and executed through the system project.

\subsection{Saving of BV Generators}
\label{sec:save_sub}
\begin{table}
    \centering
    \caption[Comparision of with and without BV generators]{\textbf{Comparision of with and without BV generators.} $GF(2^8)$: baseline model with normal generators; The others are with BV generators. A throughput of 25~Gbps is achieved with a clock frequency of 300 MHz for all three designs. }\label{tab:res_sub}
    \resizebox{\linewidth}{!}{\begin{tabular}{ccccc}
\toprule
 Type               & LUT & FF & BRAM & CLB \\
 \midrule
 $GF(2^8)$          & 61012 (7.0\%) & 88380 (5.1\%) & 152 (11.3\%) & 14006 (12.9\%)\\
 $L(2^4)$  & 56635 (6.5\%) & 72253 (4.1\%) & 96 (7.1\%) & 12009 (11.0\%) \\
 $L(2^2)$  & 53285 (6.1\%) & 72189 (4.1\%) & 96 (7.1\%) & 11740 (10.8\%) \\
 \bottomrule
    \end{tabular}}     
\end{table}

In the previous section, we demonstrated that the BV generator considerably reduces the size of a finite field multiplier. In this section, we implement three designs utilizing generator matrices with the entries taken from the finite field $GF(2^8)$, as well as with the entries taken from the BV subsets $L(2^4)$ and $L(2^2)$.  
Each of these designs is implemented with 8 load-balanced BATS CUs, where each CU is connected to an individual HBM PC through a dedicated AXI adapter. A tile size of $8\times8$ is employed for these implementations.  
We also implemented a CS-BATS code with parameters $pk=256$ and $M=16$, using a base graph with degrees $\{11, 12, 14, 14, 19, 20, 27, 32\}$.  
The encoding throughput achieved by these designs is calculated, and the resource utilization following the complete HLS implementations is compared. The results are presented in \cref{tab:res_sub}.

As observed, employing the BV generator within the complete design results in resource savings of up to 13\% in LUTs, 18\% in FFs, and 36\% in BRAM.  
Although the entire design encompasses modules beyond the compute unit, leading to slightly lower overall resource savings than initially anticipated based on the MAC unit alone, the reduction in area and propagation delay afforded by the BV generator remains significant. This is particularly advantageous for the scalability of the design, especially in applications with power and resource constraints, such as IoT device communications and wireless sensor networks.  
The achieved clock frequency shows no notable differences among the three designs, all of which comfortably meet the timing requirements. Consequently, all three designs deliver identical throughput.

\subsection{Scalablility}

\begin{table*}
    \centering
    \caption[Comparision of implementation results]{\textbf{Comparision of implementation results of various components.} All designs are implemented with BV generators in $L(2^2)$. Designs that achieve the best performance and the best balance between performance and resource utilization are highlighted. \textbf{Exp.}: experiment names; \textbf{CUs}: the number of BATS CUs; \textbf{OUT}: the number of output ports; \textbf{AXI}: the number of AXI adapters for input ports; \textbf{PC}: the number of HBM pseudo channels; \textbf{LBS}: load-balanced scheduling; \textbf{Read and Write}: the kernel-to-HBM read and write speed; \textbf{Thr.}; encoding throughput of the accelerator. }
    \label{tab:res_sca}
    \resizebox{\textwidth}{!}{\begin{tabular}{c|cccccccccccc}
\toprule
Exp.&CUs&OUT&AXI&PC&LBS       & LUT         & FF          & BRAM     & CLB           &Read&Write& Thr.\\
\hline
\textbf{A}   &\textbf{8}&1&\textbf{8}  &\textbf{8}  &\textbf{\checkmark}&\textbf{53285 (6.1\%)}&\textbf{72189 (4.1\%)}&\textbf{96 (7.1\%)}&\textbf{11740 (10.8\%)}&\textbf{12 Gbps}&\textbf{52 Gbps}&\textbf{25 Gbps}\\
B   &8&1&1  &1  &\checkmark&47998 (5.5\%)&71289 (4.1\%)&96 (7.1\%)&10692 (9.8\%) &17 Gbps& 49 Gbps& 12 Gbps\\
C   &8&1&2  &2  &\checkmark&44359 (5.1\%)&63922 (3.7\%)&96 (7.1\%)&9712 (8.9\%) &17 Gbps& 49 Gbps& 18 Gbps\\
\textbf{D}   &\textbf{8}&1&\textbf{4}  &\textbf{4}  &\textbf{\checkmark}&\textbf{45029 (5.2\%)}&\textbf{65010 (3.7\%)}&\textbf{96 (7.1\%)}&\textbf{10101 (9.3\%) }&\textbf{17 Gbps}& \textbf{52 Gbps}& \textbf{24 Gbps}\\
E   &8&1&4  &2  &\checkmark&45029 (5.2\%)&65010 (3.7\%)&96 (7.1\%)&10101 (9.3\%) &13 Gbps& 51 Gbps& 23 Gbps\\
F   &8&1&8  &8  &$\times$  &46806 (5.8\%)&67498 (3.9\%)&96 (7.1\%)&10604 (9.7\%)&9.5 Gbps& 51 Gbps&24 Gbps\\

G   &4&1&4  &4  &\checkmark&24391 (2.8\%)&38022 (2.2\%)&48 (3.6\%)&5794 (5.3\%)&12 Gbps& 36 Gbps&14 Gbps\\
H   &2&1&2  &2  &\checkmark&12841 (1.5\%)&21127 (1.2\%)&24 (1.8\%)&3171 (2.9\%)&13 Gbps& 22 Gbps&7.5 Gbps\\
I   &1&1&1  &1  &\checkmark&7203 (0.8\%)&12331 (0.7\%)&12 (0.9\%)&1790 (1.6\%)&12.5 Gbps& 56 Gbps&6.0 Gbps\\
\textbf{J}   &\textbf{8}&\textbf{2}&\textbf{8}  &\textbf{8}  &\checkmark&\textbf{49289 (5.7\%)}&\textbf{75660 (4.3\%)}&\textbf{96 (7.1\%)}&\textbf{11315 (10.4\%)}&\textbf{12 Gbps}& \textbf{36 Gbps}&\textbf{27 Gbps}\\
\bottomrule
    \end{tabular}}
\end{table*}

As discussed in previous sections, the level of parallelism and scalability in the design is largely influenced by the contention among the BATS CUs. Ideally, each CU should operate independently, as this would allow for an increase in throughput proportional to the number of CUs that can be integrated onto the chip.  
The degree of contention in our design primarily depends on the number of AXI adapters and HBM PCs. Each AXI adapter connects the input port of a BATS CU to an HBM PC. To minimize contention, the number of AXI adapters should at least match the number of HBM PCs. While multiple AXI adapters can share a single HBM PC, this setup may lead to contention among the adapters. Similarly, bundling multiple input ports of the BATS CUs into a single AXI adapter can result in contention among the CUs. It is important to note that implementing additional AXI adapters increases hardware resource consumption, whereas utilizing more HBM PCs does not incur extra hardware resource costs. However, this approach necessitates the creation and storage of individual copies of input data on each PC, thereby introducing additional memory transfer between the host and the accelerator.

Beyond the number of AXI adapters and HBM pseudo channels (PCs), the performance is also significantly constrained by the number of BATS CUs and the load-balanced scheduler.  
In this section, we analyze the impact of these four factors on the design's performance and resource utilization to determine the most balanced and scalable design configuration.  
We conducted nine experiments with different design implementations, labeled as $A, B, C, \ldots, I$. All designs were implemented using the full Vitis flow, operating at a clock frequency of 300 MHz, and profiled using Vitis Unified IDE 2023.1. The results are presented in \cref{tab:res_sca}, where we report the number and percentage of FPGA resource utilization, kernel I/O speeds, and accelerator throughput for generating 32 batches. Each design incorporates BV generators in $L(2^2)$, with a tile size of $8\times 8$. All other BATS configurations remain the same as those described in \cref{sec:save_sub}.

\subsubsection{Througput scales as the number of CUs}
Experiment~A in \cref{tab:res_sca} serves as the baseline design, featuring 8 BATS CUs and 8 individual AXI adapters for reading input data from 8 HBM PCs. This design achieves a throughput of 25 Gbps while utilizing only approximately 6\% of the LUT.  
Additionally, we implemented designs with 1, 2, and 4 BATS CUs, corresponding to Experiments I, H, and G in \cref{tab:res_sca}, respectively.

In general, we observe that resource utilization scales roughly linearly with the number of CUs when the latter increases from 2 to 8. Similarly, the throughput exhibits nearly linear scaling, rising from 7.5 Gbps with 2 BATS CUs to 25 Gbps with 8 BATS CUs.  
Interestingly, the design with 1 BATS CU achieves a throughput comparable to that of the design with 2 BATS CUs despite utilizing almost half the resources. This can be attributed to the simplification of scheduling logic and output port sharing logic (\cref{alg:data-out}) in the case of a single BATS CU, which is reflected in the significantly higher write speed observed in Experiment~I compared to Experiments H and G.

\subsubsection{Impact of port bundling}
Experiments A through D examine the relationship between resource utilization and throughput as the number of AXI adapters varies, aiming to identify the optimal balance between resource consumption and throughput performance.  
In general, an increase in the number of AXI adapters results in a higher throughput due to reduced contention, albeit at the cost of increased resource usage. However, as shown in Experiment~B, the resource consumption with just 1 AXI adapter is higher than that with 2 or 4 adapters. We hypothesize that it is because the compiler requires additional resources to manage the severe contention that arises when only a single AXI adapter is used.  
Notably, Experiment~D demonstrates that using 4 AXI adapters with 8 BATS CUs results in a throughput similar to that of 8 AXI adapters with 8 BATS CUs, but with significantly lower resource utilization. Therefore, we can conclude that sharing 1 AXI adapter between every two BATS CUs offers the best trade-off between resource utilization and throughput.

\subsubsection{Impact of HBM PC setting}
Recall that the HBM in our design features 8 crossbar switches, each containing 4 AXI connectors and 4 DDR memory controllers, resulting in a total of 32 pseudo channels. Aligned access typically yields the best performance.  
In Experiment~E, we connected 4 AXI adapters to 2 HBM PCs, with 2 adapters sharing a single PC. This configuration led to a drop in read throughput from 17 Gbps in Experiment~D to 13 Gbps, attributed to contention and unaligned access.  
Despite this, Experiment~E demonstrates that our design remains scalable and adaptable to FPGA platforms without HBM, such as those equipped with only a few DDR memory banks. The throughput decreases by only 1 Gbps when the number of PCs is halved, indicating that relatively high throughput can still be achieved on other FPGA platforms with 2 or fewer DDR memory banks.

\subsubsection{Output Port sharing}
Most of our experiments utilize a single output port with the port sharing strategy described in \cref{alg:data-out}. Since the tiles computed by the systolic array need to be transposed in a ping-pong buffer before being written to the HBM, this buffer helps prevent stalling when only a few output ports are used. Additionally, the ping-pong buffer allows full utilization of the output port width and enhances data transfer efficiency through coalescing. Consequently, our design can scale with the number of BATS CUs and maintain high throughput even with a single output port. However, as the number of CUs increases, the output port's processing capacity may eventually become saturated, leading to stalls.

In Experiment~J, we re-implemented the design from Experiment~A with 2 output ports, each connected to a separate AXI adapter. This configuration resulted in a 30\% increase in write throughput compared with the single output port design, indicating that the output port in Experiment~A was saturated to approximately 38\% of the total output throughput required in Experiment~J ($(36\times2-52) / 52 \approx 0.38$). Thus, utilizing 2 output ports achieves higher throughput than with 1 output port. Notably, despite using one additional output port, Experiment~J consumes fewer LUTs than Experiment~A due to the reduced logic resources required to manage stalling. 

\subsubsection{Load balance scheduling}
In Experiment~F, we implemented a design without load-balanced scheduling. This resulted in reduced read and write throughput due to the idling of the BATS CUs, leading to a decrease in coding throughput from 25 Gbps to 24 Gbps. Although the throughput reduction, in this case, is not substantial, the issue of unbalanced load and its impact on scheduling will likely become more pronounced with base graphs of larger degrees.

\subsection{Comparision with Software Implementation}
For comparison, we implemented the random BATS and the CS-BATS with C++ on Dell OptiPlex~7080 with an I7-10700 CPU. The compilation is performed with g++~11.4.0 on Ubuntu~22.04.3 and Linux kernel 6.8.0.  
The implementation is optimized with matrix tiling, and the finite field operation is implemented with look-up tables. 
Same as the FPGA implementation, we used $pk=256, M=16$, $GF(2^8)$, and the same base graph for the CS-BATS. We calculate the throughput of generating 32 batches using an optimized degree distribution for inactivation decoding. The average throughput achieved for 100 trials is 85 Mbps for the random BATS and 200 Mbps for the CS-BATS implementation. 

Notice that the software implementation uses only a single core without SIMD. Nonetheless, even with SIMD and other CPU-based optimization techniques, it is difficult to match the throughput, power consumption, robustness, and scalability of the hardware implementation. 

\subsection{Comparision with Previous FPGA Implementation}

\begin{table}
    \centering
    \caption{Comparision between Our implementation (Exp. D in \cref{tab:res_sca}) and the implementation in \cite{tvt}. The number of BRAMs in our implementation is normalized from 36K BRAM to 10K BRAM for a fair comparison.}
    \label{tab:compare_tvt}
    \begin{tabular}{c|ccccc}
    \toprule
         & LUT & FF & BRAM & Freq. & Throughput\\
    \midrule
        Ours & 45,029 & 65,010 & 345 & 300 Hz & 24 Gbps\\
        \cite{tvt} & 49,402 & 68,787 & 178 & 106 Hz & 1.2 Gbps\\
    \bottomrule
    \end{tabular}
\end{table}

In \cref{tab:compare_tvt}, we compare our implementation (Exp. D in \cref{tab:res_sca}), which has the best resource-performance trade-off, with the implementation in \cite{tvt}. Since \cite{tvt} is implemented on Intel Cyclone V FPGA, while ours is based on AMD U50 FPGA, we normalize the number of BRAMs for a fair comparison. The Intel Cyclone V uses 10Kb BRAMs, while the AMD U50 uses 36Kb BRAMs. Therefore, we convert the original number of BRAMs used in our implementation to the corresponding number needed with 10Kb BRAMs (the original number of BRAMs times 36 bits, then divided by 10 bits: $96*36/10\approx 345$).
As both boards mainly use the 6-input LUTs, we compare the number of LUTs and the number of FFs directly.
We can observe that with fewer LUTs and FFs, our implementation achieves much higher throughput than the previous one, even if we take the difference in clock frequency due to the hardware into account. 
In order to resolve the contentions among the CUs and provide more parallelism, more BRAMs are used as buffers in our design. However, this comparison is only for reference, as we convert the number of 36Kb BRAMs to the number of 10Kb BRAMs according to the number of bits available in the memory, but some BRAMs may not be fully utilized, which leads to over-estimation during this conversion. 

\section{Conclusion and Discussion}
This work introduced a novel network coding accelerator based on the Alveo U50 FPGA. We investigated the integration of the modern FPGA with HBM and the state-of-the-art network coding algorithm, the BATS code. We took an algorithm-hardware co-designing approach, where a hardware-friendly variant of the BATS code, the CS-BATS, was proposed for use. We first discussed the design of the basic unit of our accelerator, the BATS CU, which achieves both batch-level parallelism and tile-level parallelism. Then, we investigated the I/O contention and the solutions we proposed to resolve the contention, such as tiling-and-data-access decoupling, output port sharing, and AXI-adapter-HBM co-designing. 
To reduce the computational complexity of finite field operations, we proposed the BV generator that exploited the characteristic of the generator matrix to reduce the size of the finite field multiplier by an order of magnitude. 
We analyze the BV generators theoretically and experimentally, guaranteeing a negligible effect on the coding performance of the BATS code.
By implementing the design, we showed that the BV generator saves up to 36\% of hardware resources for the complete design. 
We also demonstrated that with small hardware resources,
we achieved a throughput of 27 Gbps, outperforming an optimized software implementation on an I7-10700 CPU significantly, which has a throughput of around 85~Mbps. 
With an extensive ablation study, we analyzed the impact of different components and concluded with the best design choice that achieves a sweet spot between resource utilization and accelerator throughput. 


\bibliographystyle{IEEEtran}
\bibliography{IEEEabrv, paper.bib}

\end{document}